# Hybrid Graphene-Plasmon Gratings

Tianjing Guo[1] and Christos Argyropoulos[2,*]

[1]Institute of Space Science and Technology, Nanchang University, Nanchang 330031, China
[2]Department of Electrical Engineering, The Pennsylvania State University, University Park, PA 16803, USA
*cfa5361@psu.edu

**Abstract:** Graphene can support surface plasmons with higher confinement, lower propagation loss, and substantially more tunable response compared to usual metal-based plasmonic structures. Interestingly, plasmons in graphene can strongly couple with nanostructures and gratings placed in its vicinity to form new hybrid systems that can provide a platform to investigate more complicated plasmonic phenomena. In this Perspective, an analysis on the excitation of highly confined graphene plasmons and their strong coupling with metallic or dielectric gratings is performed. We emphasize the flexibility in the efficient control of light-matter interaction by these new hybrid systems, benefiting from the interplay between graphene plasmons and other external resonant modes. The hybrid graphene-plasmon grating systems offer unique tunable plasmonic resonances with enhanced field distributions. They exhibit a novel route to realize practical emerging applications, including nonreciprocal devices, plasmonic switches, perfect absorbers, nonlinear structures, photodetectors, and optical sensors.

## I. INTRODUCTION

Graphene is the flagship two-dimensional (2D) material made of carbon atoms arranged in a hexagonal lattice. It has attracted a surge of research interest for its unique electrical, mechanical, optical, and thermal properties.[1–4] It exhibits distinct advantages over metals leading to the realization of plasmonic effects at infrared (IR) and terahertz (THz) frequencies, where it can support surface plasmons (SPs) with high confinement and relatively low loss.[5] Graphene plasmons (GPs) can become highly tunable by applying electrostatic gating voltage or chemical doping,[6,7] [8,9]making graphene an important building block of envisioned reconfigurable devices. Another important advantage of GPs is their strong light confinement that becomes even stronger by further increasing the doping concentration in graphene.[10,11] Moreover, GPs demonstrate much lower loss compared to noble metal plasmons mainly due to graphene's higher carrier mobility, especially in the case of high doping.[10,12] As a result, many theoretical and experimental studies have focused on demonstrating GPs due to their potential applications in various optoelectronic emerging technologies.[13–17]

Graphene monolayers can be tailored into various geometries, such as ribbons,[18–20] disks,[21,22] and crosses,[23] with the goal to improve the GPs excitation. However, the process of graphene patterning to various shapes is challenging to be experimentally implemented and can cause detrimental effects on graphene's response. Another more practical approach to excite GPs is to integrate graphene monolayers with various already resonating structures, including plasmonic nanoantennas, gratings, and metamaterials, forming new hybrid plasmonic or dielectric systems. These hybrid graphene-based structures have the potential to rapidly increase the interaction strength between graphene and incident light radiation, leading to a variety of new IR and optical devices with exciting functionalities, such as tunable absorbers,[24,25] efficient modulators,[26,27] and photodetectors.[28,29] As an example, optical gratings can provide strong localized electric fields and can be used to excite even more confined and enhanced GPs due to the formation of hybrid graphene-plasmon gratings (HGPGs).[12,27,30–33] The hybrid designs consisting of gratings combined with graphene benefit from the merits of both systems. More specifically, the excited plasmons in graphene can strongly couple with the grating structure due to their proximity and coupling of their near-fields,[34,35] usually causing enhanced resonant absorption that is a desirable outcome for most applications. In addition, metallic gratings used in such hybrid structures can work as gate electrodes to further tune the resonant frequency of the designed graphene-based device.

The combination of graphene and metallic (plasmonic) nanostructures offers a unique way to enhance absorption at their resonance, where sharp peaks of reflectance or transmittance, combined with enhanced fields are present. This is the fundamental response behind the realization of a wide range of potential applications, including sensing, light modulation, and photodetection.[36–41] In addition, these hybrid devices can have a tunable operation at mid-infrared (IR), far-IR, and THz frequencies just by using different grating periods. As a potentially useful application, HGPGs can be used to detect the presence of molecules or ions in a solution by measuring changes in the plasmonic properties of the total hybrid system.[42,43] This approach can be used for the sensitive detection of DNA, proteins, and other biomolecules. Interestingly, graphene does not play the leading role in this type of sensing response. However, it provides the required tunability and biocompatibility, making it an ideal building block for new sensing and photodetecting devices. The superior reconfigurability guarantees that HGPGs can work as selective sensors for diverse applications, since their sensing frequency can be tuned in a wide range. HGPGs can also be used to create tunable optical filters and switches.[44–46] The plasmonic properties of HGPG devices can be modulated by applying an external voltage, allowing for precise control over the transmission and reflection of light. Plasmonic absorption is also beneficial to graphene-based photodetectors that have been demonstrated to achieve ultrafast photodetection speeds combined with broadband operation.[47–52] The photodetector responsivity is enhanced by utilizing HGPG designs simply because they capture light more efficiently.[53,54] With the additional advent of various relevant nanofabrication techniques, HGPGs are expected to lead to new exciting applications emerging from these novel configurations. In this Perspective, we provide a detailed overview of the recent advances of various HGPGs structures and their potential applications.

## II. STRONG COUPLING OF GRAPHENE-PLASMON GRATINGS

The use of metallic subwavelength gratings is an appealing approach to boost the excitation efficiency of GPs along a continuous graphene monolayer. By placing graphene below a metallic aperture, a novel class of plasmon resonance was reported, where strong coupling occurs between the newly created hybrid plasmon mode and incident THz radiation.[45] The hybrid graphene plasmonic structure is shown in Fig. 1(a). It is composed of a metallic grating patterned on top of a continuous graphene monolayer. Higher order modes present at longer frequencies are not considered and only the fundamental plasmon mode is investigated in this study. It is found that only when the metal contact is much wider than the grating channel $w$, a strong resonance can occur combined with high absorption with spectrum demonstrated in Fig. 1(c). Meanwhile, a maximum absorption value of 50% is reached, implying very high field confinement at this resonant plasmon frequency. On the contrary, Fig. 1(d) depicts a far lower absorption response for the complementary structure shown in Fig. 1(b), i.e., periodic graphene ribbons without metallic grating. To interpret the above absorption results, the charge density for the two cases is calculated at the resonant frequency and demonstrated by the three-dimensional (3D) distributions in Figs. 1(e) and 1(f), respectively. The former case exhibits a resonant circuit behavior that interacts strongly with the incident wave leading to high absorption, where the metal contacts act as capacitors and the graphene as inductors. While the latter work consists only of inductors (graphene) and exhibits much weaker resonant response. Hence, the incorporation of plasmonic gratings into graphene can substantially improve the THz absorption performance.

An alternative relevant design was recently demonstrated, made of a continuous, large-area graphene sheet grown on silica substrate and coated with a square periodic array of metallic nanoparticles.[55] Figure 1(g) presents this configuration, where the resulted diffractive coupling depends on the grating period $\Lambda$. Hence, the resonant frequency of the induced grating-coupled graphene plasmon can be tuned by the grating period $\Lambda$ and graphene carrier density, as illustrated in Fig. 1(h). The resonant frequencies of graphene plasmons are tuned across the entire THz regime by using different micrometer-scale array periods $\Lambda$. The measured transmission spectra are demonstrated in Fig. 1(i), when the hybrid device has period $\Lambda = 1.5\,\text{um}$, where a transmission dip is observed that becomes more pronounced for higher applied gate voltage ($V_{GS}$) values. However, the resonant frequency is always constant and approximately equal to 7 THz. This work illustrated that the plasmonic transmission and absorption behavior is not consistent with the predicted theoretical analysis when large variations in carrier density N across the graphene sample are present.[55] This useful result stressed the importance of uniform carrier distribution along graphene to realize strongly tunable plasmonic devices.

The transformation of electromagnetic waves into propagating plasmons in graphene was also realized by the asymmetric dual-grating gate configuration operating in reflection mode depicted in Fig. 1(j).[56] In this work, the wide-area graphene monolayer is separated from the dual-grating gate by a thin dielectric sheet. The interaction of the incident THz wave with graphene is greatly enhanced by canceling the transmission and weakening the reflection due to the metallic substrate and Fabry-Perot resonances, respectively. This structure transforms normal incident THz waves into graphene propagating plasmons. The efficiency of the transformation is characterized by the so-called transformation coefficient, which can have negative or positive values when the graphene plasmon propagates in negative or positive directions, respectively. The absorbance A and transformation coefficient $T_p$ are presented in Figs. 1(k) and 1(l), respectively, by optimizing the grating's asymmetry and fixing the narrower gate width. When the relaxation time of graphene is $\tau = 1\,\text{ps}$, the transformation coefficient can reach its maximum value of 51% at 4 THz. Moreover, the absorbance shown in Fig. 1(k) is very close to 100% at the maximum transformation coefficient point, mainly due to the excited Fabry-Perot resonances. This work demonstrates the possibility to realize efficient transformers of the incident THz waves into propagating plasmons along graphene.

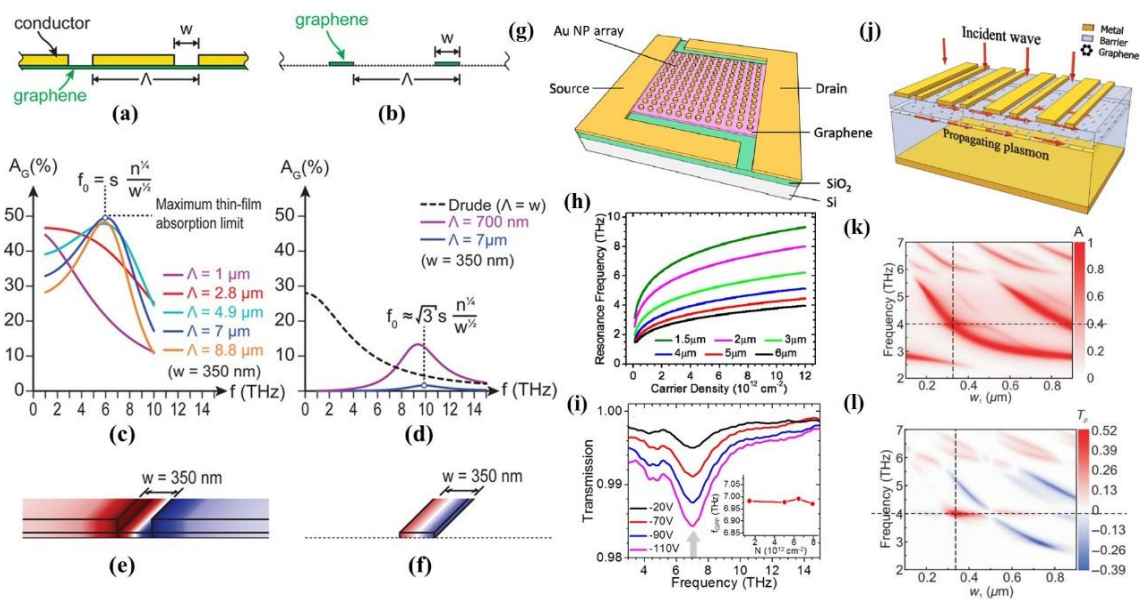

Figure 1. Tunable hybrid metal-graphene plasmons.
(a) Hybrid metal-graphene structure to achieve strong coupling between the plasmon mode and incident THz wave. (b) Isolated graphene ribbons leading to the complementary structure of (a). (c)-(d) Absorption spectra of the structures presented in (a) and (b), respectively. (e)-(f) Charge density profiles along the structures shown in (a) and (b), respectively.[45] Reproduced with permission from Jadidi *et al.*, Nano Lett. 15, 7099 (2015). Copyright 2015, American Chemical Society. (g) Hybrid graphene structure with continuous large-area graphene sheet coated by a periodic array of gold nanoparticles. (h) First-order resonant frequency as a function of carrier density for different period values $\Lambda$. (i) Measured transmission spectra of the structure in (g) for different gate voltage values $V_{GS}$. The inset is the minimum transmission frequency as a function of carrier density N.[55] Reproduced with permission from Tantiwanichapan *et al.*, ACS Photonics 4, 2011 (2017). Copyright 2017, American Chemical Society. (j) Dual-grating gated graphene configuration to excite propagating GPs. (k) Absorption as a function of frequency and width $w_1$ of the wider gate finger. (l) Transformation coefficient as a function of frequency and width $w_1$.[56] Reproduced with permission from Fateev. *et al.*, Phys. Rev. Appl. 11, 064002 (2019). Copyright 2019, American Physical Society.

Hence, strong coupling phenomena provide a unique route to efficiently excite plasmonic resonances in graphene and graphene-based nanostructures that can be used in various emerging applications. Phonons generated at various materials can also couple to graphene plasmons when strong coupling is achieved.[57] Multiple plasmonic resonances were observed experimentally in graphene nanoribbons leading to four absorption peaks arising from the strong coupling of GPs and three surface optical phonon modes.[58] Additionally, direct experimental evidence was provided that the coupling of a graphene/SiC support with gallium nanoparticles-localized surface plasmon resonances can be controlled over a broad frequency range, contributing to the enhancement of Raman modes.[59] Plasmonic strong coupling effects were also investigated experimentally with graphene nanoribbon structures,[60] enabling field-enhanced optical spectroscopy,[61] biological sensing,[62] and complete absorption.[24,35]

Interestingly, the efficient coupling of GPs to resonant modes excited by metamaterial structures was achieved with hybrid graphene-metamaterial systems.[63] More specifically, complete modulation in the strength of the metamaterial electromagnetically induced transparency was realized because of the near-field coupling between monolayer graphene and THz resonant metamaterials.[64] Furthermore, the incident photons were efficiently coupled to plasmons along the graphene layer by constructing a doubly resonant plasmonic configuration, leading to strong optical absorption with a large tuning range.[65] Strong coupling effects were also reported between GPs and special evanescent wave modes induced by near-field perturbations between dielectric and metallic gratings.[66] Moreover, tunable large THz modulation was demonstrated by coupling a gated continuous graphene layer to a non-resonant subwavelength scale meta-atom structure.[67] Finally, effective directional couplers were designed by combining multilayer graphene-based cylindrical waveguides and metal-based waveguides.[68]

The excitation of magnetic polaritons in metal gratings with deep corrugations and their coupling with plasmons in graphene was accomplished based on the hybrid plasmonic system shown in Fig. 2(a).[35] The strong coupling effect demonstrated in this work resulted in substantial enhancement to the system's absorption by boosting the intrinsic plasmonic resonance of graphene. The absorption enhancement was utilized indirectly to determine the coupling effect strength in this work. The hybrid grating in Fig. 2(a) generates a highly localized magnetic field inside its trenches forming magnetic polaritons (MPs), also known in the literature as spoof plasmons.[69,70] Figure 2(b) shows the absorptance spectrum for bare and graphene-covered grating structures. The obtained absorptance is only 0.35 for the bare grating at the MP resonance frequency of 1041 cm$^{-1}$. However, the absorptance is substantially boosted to 0.94 after covering the grating with graphene, while its resonant frequency exhibits a blueshift. The absorptance enhancement and accompanied resonance shift clearly indicate the strong coupling effect between the metallic grating and graphene. This work also demonstrated the effect on the coupling strength when graphene is elevated above the grating surface. As expected, the coupling effect deteriorates when the graphene sheet detaches from the grating and its distance is increased. In addition, the coupling is modified by the geometrical parameters and graphene chemical potential values. Figure 2(c) depicts the absorptance spectra under different chemical potentials. The coupling strength is highest when $\mu=0.3$ eV and deteriorates when $\mu$ slightly changes to 0.28 eV or 0.32 eV.

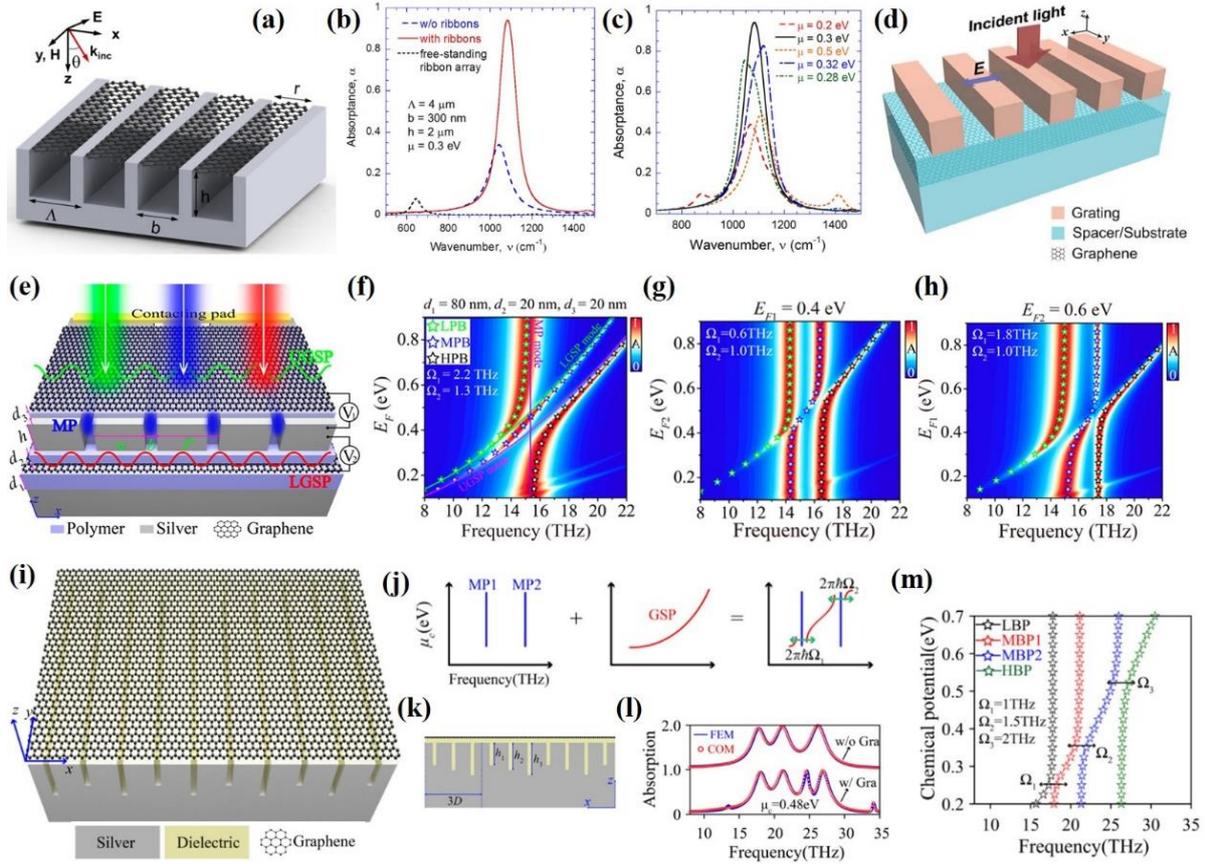

Figure 2. Strong coupling in grating-graphene plasmonic hybrid systems.
(a) Geometry of graphene ribbon-grating hybrid structure. (b) Absorption spectra of grating without graphene, hybrid graphene-grating structure, and bare graphene ribbon array. (c) Absorption spectra of the graphene ribbon-grating hybrid structure under different graphene Fermi levels.[35] Reproduced with permission from Zhao *et al*., ACS Photonics 2, 1611 (2015). Copyright 2015, American Chemical Society. (d) Schematic of grating-assisted double-layer graphene hybrid system.[71] Reproduced with permission from Zhao *et al*., Opt. Lett. 41, 5470 (2016). Copyright 2016, Optica. (e) Geometry of grating-assisted double layer graphene hybrid structure consisting of a silver grating placed between two graphene layers. (f) Absorption plots of the hybrid system as a function of frequency and graphene chemical potential $E_F$. Here, the upper- and lower-layer graphene layers have the same chemical potential. (g) Absorption plots as a function of frequency and lower-layer graphene chemical potential $E_{F2}$ when the upper-layer graphene chemical potential is fixed to $E_{F1} = 0.4$ eV. (h) Absorption plots as a function of frequency and upper-layer graphene chemical potential $E_{F1}$ when the lower-layer graphene chemical potential is fixed to $E_{F2} = 0.6$ eV.[72] Reproduced with permission from Qing *et al*., ACS Photonics 6, 2884 (2019). Copyright 2019, American Chemical Society. (i) Compound silver grating covered by a continuous graphene layer. (j) Interaction sketch among graphene plasmon and two different MP modes. (k) Geometry of the resulted hybrid system when three different MP modes are excited. (l) Simulated and theoretically calculated absorption spectra of the structure in (k) for the cases with and without graphene cover. (m) Dispersion relation of the resulted four hybrid modes.[73] Reproduced with permission from Qing *et al*., Carbon N. Y. 145, 596 (2019). Copyright 2019, Elsevier Ltd.

The dielectric diffractive grating in Fig. 2(d) can also be used to efficiently couple the incident field and excite plasmonic resonances along a double graphene layer.[71] Strong coupling is demonstrated by changing the graphene layer separation distance, substantially modifying the transparency window of the entire hybrid system. Similarly, an alternative hybrid system is demonstrated in Fig. 2(e) that consists of a metallic grating separating two continuous graphene films.[72] Three different modes are excited in this hybrid structure, i.e., the upper-layer graphene plasmon (UGP), lower-layer graphene plasmon (LGP), and MP. Classic coupled oscillator models are used to analyze the coupling behavior of these modes, which leads to two anti-crossing responses characterized by different Rabi splitting values and three new hybrid-polariton modes. The strong coupling between MP and the two graphene plasmon modes is explored by calculating the absorption contour plots of this hybrid system. Spectral splitting is observed in the absorption spectra due to the strong coupling between the MP and UGP modes. The obtained anti-crossing behavior occurs with a moderate splitting energy. The MP mode is coupled to the two graphene plasmon modes simultaneously leading to a strong interaction between UGP (or LGP) and MP modes that gives rise to hybrid dispersion bands. Figure 2(f) illustrates the absorption contour plot as a function of frequency and graphene chemical potential $E_F$ when upper- and lower-layer graphene have the same chemical potential ($E_F = E_{F1} = E_{F2}$). Strong coupling exists between the MP and UGP (or LGP) modes as $E_F$ increases, leading to the formation of three distinct hybrid dispersion bands. The spacer thickness also plays an important role on the coupling between the MP and UGP or LGP modes. More importantly, the coupling phenomena can be dynamically tuned by controlling the chemical potential of each graphene layer separately, as depicted in Figs. 2(g) and 2(h), where the doping is fixed to one graphene layer and vary in the other. This study presented an efficient platform to explore the coupling effect achieved by hybrid graphene-based systems.

The aforementioned multimode coupling provides additional energy exchange channels and is expected to attract increased attention compared to the single coupling mode. Towards this end, the hybrid coupling among GP and MP modes was theoretically investigated by the multi-coupled hybrid system shown in Fig. 2(i), consisting of a thin dielectric spacer layer, asymmetric silver grating filled with dielectric, and a graphene monolayer placed on top.[73] Based on the different trench depths of the compound grating, two distinct MP modes (MP1 and MP2) are excited by this composite configuration. In addition, the hybrid system is dynamically tuned by controlling the chemical potential of graphene. Graphene plasmon and MP modes are excited simultaneously along a practical design of this hybrid structure. The multimode interaction characteristics are depicted in Fig. 2(j). Rabi splitting occurs when the graphene plasmon is coupled to MP1/MP2 modes, generating two new hybrid plasmonic modes characterized by three dispersion bands. Moreover, an additional MP mode is introduced in the hybrid system by designing one periodic trench depth to be deeper compared to other trenches in the silver grating, as shown in Fig. 2(k). Based on the coupled oscillator model theoretical calculations and finite element method (FEM) simulations, the obtained absorption spectra are presented in Fig. 2(l) for the two cases with and without graphene monolayer. Three absorption peaks are observed when graphene is absent, corresponding to the MP1, MP2, and MP3 modes, respectively. When graphene is included in the system, the third absorption peak splits into two peaks because of the hybridization of the MP3 mode with the second order graphene plasmon mode. Figure 2(m) displays the dispersion relation of the resulted four hybrid modes. Interestingly, all four dispersion bands exhibit an anti-crossing strong coupling response. Hence, this type of complicated multimode hybrid system coupled with graphene will further enhanced light-matter interaction.

## III. HYBRID GRAPHENE-PLASMON GRATINGS APPLICATIONS

### A. Absorption enhancement

Graphene monolayers exhibit only 2.3% absorption in optical frequencies mainly due to their single-atom ultrathin thickness.[6,26] Recently, researchers have made excellent progress to boost the light absorption in graphene.[74] Towards this goal, microcavities were used to enhance absorption by making electromagnetic waves propagate through graphene multiple times.[75] Alternatively, the overall optical path length was increased along graphene by draping it over optical waveguides leading to increased overall absorption.[76] Moreover, perfect absorption was also obtained with hybrid graphene-metal systems when strong coupling and other interference effects exist between the THz graphene plasmons and metallic grating resonances.[30] Consequently, coherent perfect absorption was reported by patterning graphene monolayers into periodic arrays of patch resonators placed on each side of an ultrathin dielectric substrate, i.e., constructing an asymmetric bifacial graphene metasurface.[24] When patterned graphene[13,21] was placed on a metallic or dielectric substrate, perfect graphene absorbers were realized by satisfying some specific constraints.[21,36,77–79] Prominent absorption peaks were achieved by using gated graphene nanoribbons, benefiting from localized plasmon excitations excited along their surfaces.[13] It was also reported that perfect THz absorption can be attained based on patterned graphene, however, with relatively low tunability.[80] In addition, anisotropic graphene plasmonic structures were explored in the mid-IR range, where tunable spectral selectivity was accomplished with localized plasmonic resonances.[81] Finally, tunable plasmon-induced absorption was realized based on graphene-assisted metal-dielectric gratings.[82,83]

As mentioned before, graphene monolayers can be coupled to various nanostructures to form new hybrid plasmonic systems. Strong enhancement of optical absorption was reported by a graphene-insulator-metal hybrid plasmonic device. This coupling was induced by the increased diffraction efficiency of the metallic (gold) grating combined with resonant coupling between plasmons generated along the grating and graphene hybrid system.[65] Additionally, plasmonic gratings covered by graphene sheets were proposed to enhance absorption due to the strong localized electric field from the induced magnetic resonances at the grating.[35,84] Alternatively, it was experimentally demonstrated that highly confined propagating surface plasmon polaritons (SPPs) can be excited in monolayer graphene at IR frequencies by using a silicon diffractive grating combined with graphene.[34]

An example of a metal/dielectric/metal grating structure covered by graphene is demonstrated in Fig. 3(a).[33] The system is modeled by full-wave simulations, as well as theoretical analysis based on the equivalent circuit model where the dielectric layer is treated as a capacitor while the metal and graphene are considered inductors. The near-IR absorption spectrum is calculated for three different systems: hybrid graphene-covered grating, grating without graphene, and bare graphene monolayer. The absorptivity of the hybrid graphene-covered grating structure is depicted in Fig. 3(b) and can reach up to 0.9 values, much higher than the other two configurations. The physical mechanism behind these results is the strong coupling between MP modes excited at the metal/dielectric grating structure and graphene. The sharp Fano resonance achieved by these hybrid structures can have potential applications in improving near-IR light absorption.[38]

The hybrid structure shown in Fig. 3(c) is composed of a metal grating coupled to a graphene/waveguide system. This configuration operates at far-IR where a sharp asymmetric resonance line profile is achieved arising from the broad graphene plasmon resonance that interferes with the sharp planar waveguide resonance mode. The asymmetric profile of the resulted Fano resonance is tuned to symmetric by changing the mobility of graphene from 0.5 $m^2$/Vs to 2 $m^2$/Vs leading to tunable light absorption. Graphene absorption decreases with the increase in graphene's mobility since lower graphene mobility always leads to higher material loss. The absorption efficiency can reach 0.76 when graphene mobility $\mu = 0.5$ $m^2$/Vs is used, as demonstrated in Fig. 3(d). The almost linear relationship between the resonant wavelength and surrounding refractive index of this hybrid system is shown in Fig. 3(e) to demonstrate the structure's sensitivity. The "right" resonance refers to longer wavelengths while the "left" refers to shorter ones. The sensitivity can reach 5.21 um/RIU for the "right" resonance, much higher than 0.242 um/RIU obtained from the "left" one. Hence, this hybrid structure can have potential applications as self-calibrated refractive index sensor.

Composite metallic gratings covered by a graphene layer are demonstrated in Fig. 3(f) that can lead to wider absorption bandwidth and increased absorption efficiency.[85] The composite system (bottom) is based on two regular gratings with optimal dimensions and small (top) or large (middle) periods. Not only exhibits high absorption but also achieves wide spectral bandwidth. Figure 3(g) displays the absorption spectra for transverse magnetic (TM) polarized waves under different scenarios. The absorption efficiency is enhanced in the case of compound grating covered by graphene compared to the scenario without graphene. Importantly, the compound grating can absorb 90% of incident light at the resonance wavelength of 1.67 um with a spectral bandwidth around 0.3 um. However, the absorption is increased to 98% after being covered by graphene at the same resonant frequency, while the spectral bandwidth is doubled to approximately 0.6 um.[85] Furthermore, the absorption is computed as a function of operation wavelength and incident angles with results shown in Fig. 3(h) for the cases of compound grating with and without graphene. The computed absorption is found to be insensitive for a broad range of incident angles from 0° to 80°. It also exhibits wider bandwidth when graphene is used compared to the no graphene case. The electric field and Poynting vector distributions are plotted in Fig. 3(i) at off-resonance (top figure) and on-resonance (middle figure) to better understand the absorption physical mechanism. The electric fields are trapped among the two grating trenches at resonance, as shown in the middle caption in Fig. 3(i), leading to higher absorption. However, the electric fields are confined just in the middle trenches along their openings at off-resonance, as depicted in the top caption in Fig. 3(i). Hence, the absorption is much lower for the off-resonance case compared to on-resonance. In addition, the energy flow diagram demonstrated in the bottom caption in Fig. 3(i) also illustrates this effect, i.e., the incoming energy is concentrated on the interface between graphene and the top edges of the trench opening. This absorber design exhibits strong absorptance combined with wide spectral bandwidth and omnidirectional response. It can have applications in the design of new photodetector and energy harvesting systems.

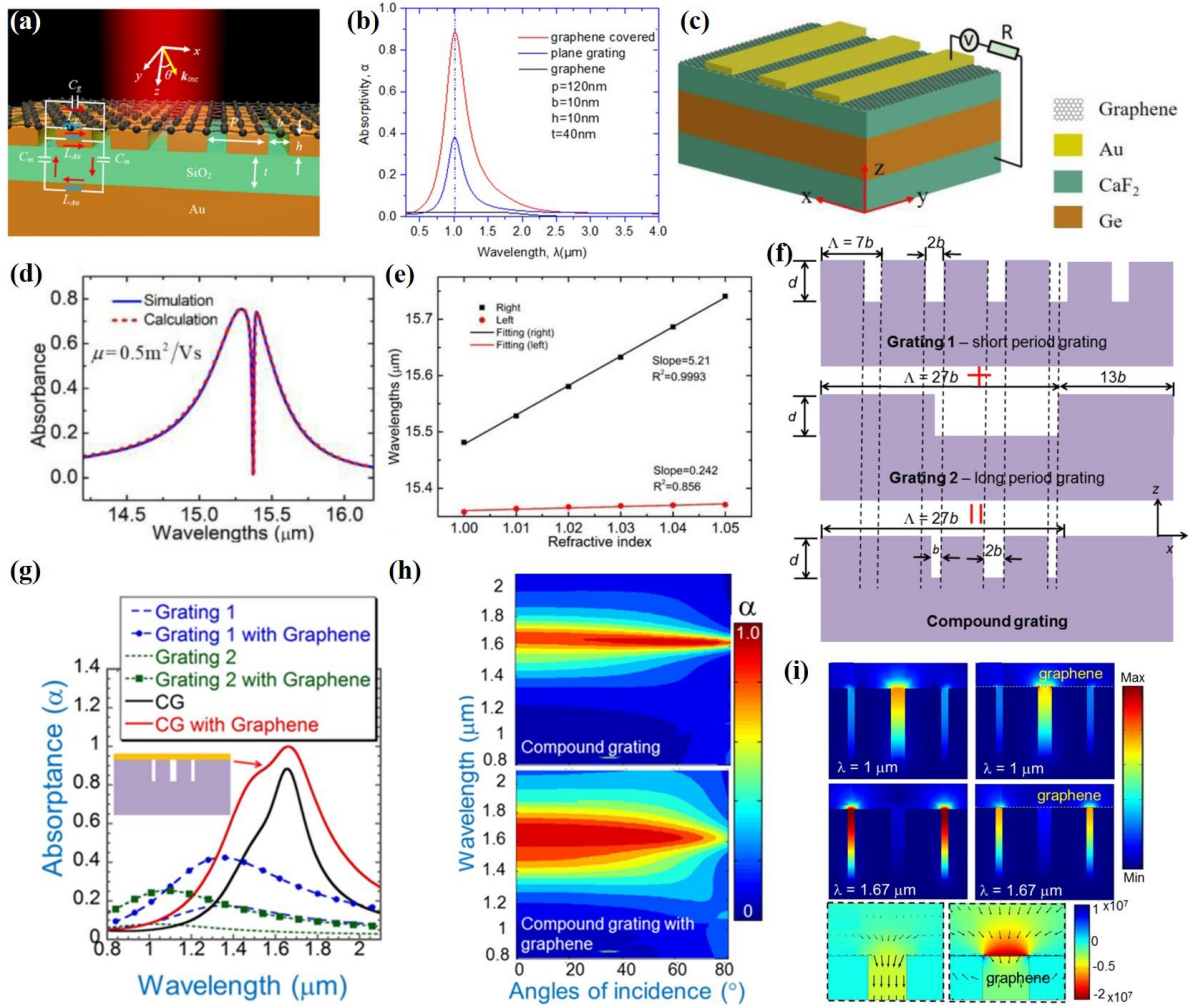

Figure 3. Hybrid graphene-grating structures to achieve enhanced absorption (theoretical results). (a) Hybrid system with graphene covering a metal/dielectric/metal grating design. The effective circuit model is used to verify the simulation results. (b) Absorption for different systems including hybrid structure, grating, and bare graphene.[33] Reproduced with permission from Pan *et al.*, Opt. Express 25, 16400 (2017). Copyright 2017, Optica. (c) Hybrid structure composed of metal grating, monolayer graphene, and planar waveguide. (d) Simulated and theoretically calculated absorption of the designed hybrid structure when graphene mobility is fixed to 0.5 m$^2$/Vs. (e) Relationship between the resonant operating wavelength and surrounding refractive index.[38] Reproduced with permission from Ruan *et al.*, J. Light. Technol. 39, 5657 (2021). Copyright 2021, IEEE Xplore. (f) Construction of compound grating by superimposing two short and long period grating designs. (g) Absorption spectra of short period, long period, and compound grating with and without graphene layer cover. (h) Absorption of compound gratings with (bottom) and without (top) graphene layer cover as a function of operation wavelength and incident angles. (i) Electric field and Poynting vector distributions in one period of the compound grating. Top and middle figures show the electric fields at off- and on-resonance wavelength, respectively. The bottom figures demonstrate the Poynting vector at resonant wavelength.[85] Reproduced with permission from Nguyen-Huu *et al.*, IEEE J. Sel. Top. Quantum Electron. 27, 1 (2021). Copyright 2021, IEEE Xplore.

Besides numerical simulations of HGPGs, experimental realization of broadband THz absorbers based on graphene was recently reported,[86] where an effective absorption bandwidth of ⩾90% ranging from 1.54 to 2.23 THz was achieved but with a relatively low electrical tunability. In another related work, the absorption was greatly boosted by placing a graphene monolayer into a Salisbury-screen-type absorber configuration.[87,88] This type of structure can provide enhanced fields in the vicinity of the graphene layer. The hybrid graphene Salisbury screen composed of graphene plasmonic resonators placed a quarter wavelength away from the back reflector was reported to absorb almost 25% of incident light, more than 10 times larger than that of monolayer graphene in the same frequency range.[89] Furthermore, electrically tunable perfect absorbers were reported experimentally by integrating metallic gratings into similar graphene Salisbury screens.[32] The resulted hybrid structures were composed of a metal reflector, polyimide layer, periodic aluminum grating, and monolayer graphene covering the entire design, as depicted in Fig. 4(a). The graphene Fermi level ($E_F$) was varied by tuning the externally applied gate voltage ($V_g$) in this configuration. The interaction between graphene and incident THz waves was further enhanced by a metallic grating embedded into the hybrid absorber design. Figure 4(b) illustrates the computed absorbance of two absorber designs with and without the grating. Both can achieve perfect absorption but the design without the metallic grating requires very high Fermi level values ($E_F = 1.05$ eV), which is hard to experimentally realize. However, the absorber design with grating can achieve perfect absorption for low Fermi level values ($E_F = 0.3$ eV) due to the improved enhancement in the interaction between graphene and incident electric field induced by the metallic grating. The measured reflection is presented in Fig. 4(c) under normal incidence for different gating voltage values and frequencies. The reflection can reach zero values around 0.72 THz, where perfect absorption is obtained, as presented by the green line in Fig. 4(d). The electrical modulation depth is also calculated at the resonance frequency and the maximum modulation depth approaches 29% that is shown by the blue line in Fig. 4(d). Hence, these results reveal potential applications of these hybrid absorber designs in THz modulators and switches.

Furthermore, the experimental demonstration of peak absorption over 99% was reported for a hybrid structure composed of a monolayer graphene sandwiched between a 1D polymethyl-methacrylate (PMMA) grating and a silica layer,[90] as shown in Fig. 4(e). The optical and scanning electron microscope (SEM) images of the fabricated structure are presented in Figs. 4(f) and 4(g), respectively. It is found that we can distinguish samples with graphene from those without graphene by inspecting the obtained optical images. The fabricated structure is measured with confocal microscope setup, which is schematically illustrated in Fig. 4(h). The measured absorption spectra of the proposed absorber design with different grating periods for transverse electronic (TE) polarization are shown in Fig. 4(i), where peak absorption of 99.6%, 99.1% and 98.6% are achieved with grating period of 1230, 1254 and 1270 nm, respectively. Importantly, it is observed that the measured results are in excellent agreement with the simulation results, confirming complete optical absorption of graphene-based hybrid structures in the near-IR range. Note that the mid-IR spectral range is equally important because of many applications that include IR vibrational spectroscopy, sensing, and thermal radiation control. Electronically tunable perfect absorption is experimentally demonstrated with the HGPG shown in Fig. 4(j).[91] In this type of structure, the metallic slit is 100 nm wide while the 50 nm wide graphene ribbon is off to one side. The narrower slit in this configuration can more efficiently confine the incident radiation, leading to larger field enhancement and much higher light absorption compared to the configuration where the graphene ribbon is located in the center of the metallic slit. Figure 4(l) presents the tunable absorption map of the proposed HGPG design, as a function of frequency and graphene Fermi level. By performing experimental measurements, a peak absorption of 96.9 % is obtained in the hybrid structure at 1389 cm$^{-1}$, revealing the potential of far-reaching applications in graphene-based active IR optical components. To assess the absorption level of HGPG designs, we compared the peak absorption and the operation range obtained with numerical simulations and experimental measurements, as seen in Table 1. It is observed that nearly perfect absorption can be obtained by hybrid graphene-grating structures from IR to THz regime. However, most structure designs can achieve only narrowband response or single-frequency operation. Future efforts will be dedicated to obtaining broadband absorption performance. In addition, it is still challenging to shift the operation wavelength to near-IR or even visible frequency bands, mainly due to the fabrication difficulties in addition to graphene property limitations.

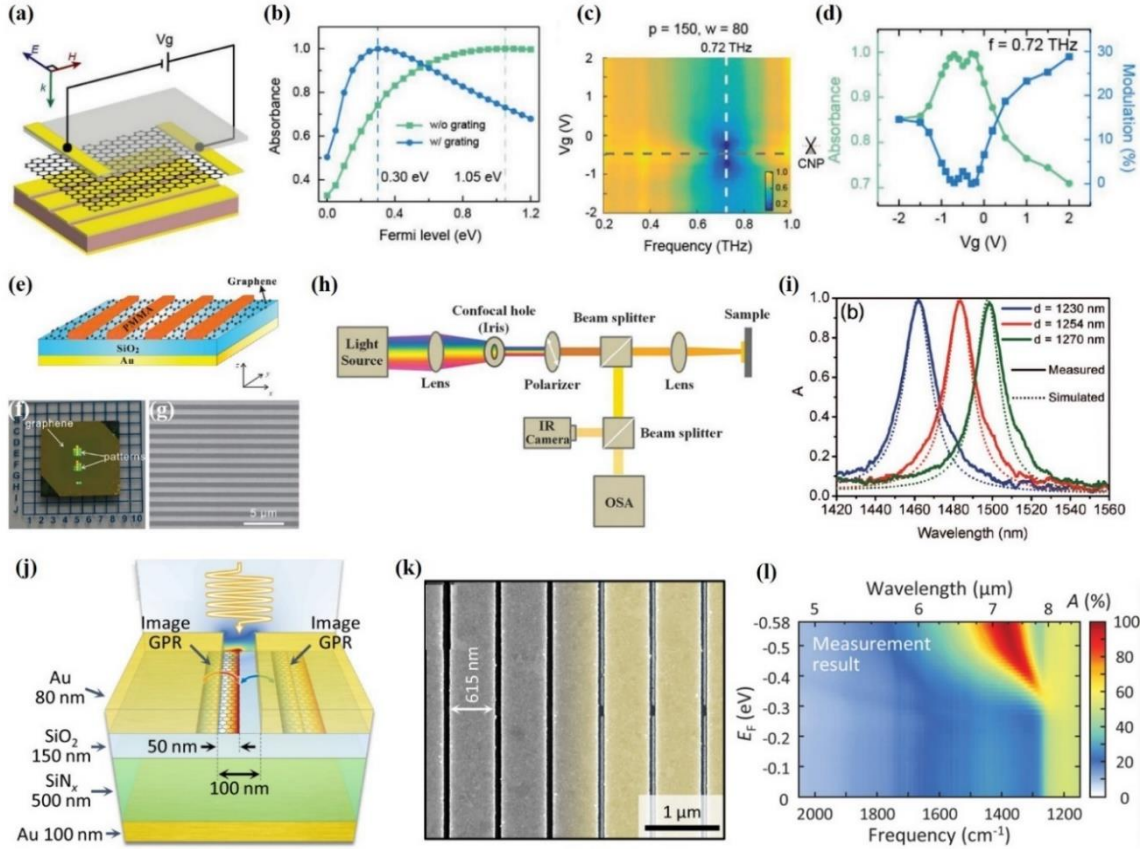

Figure 4. Hybrid graphene-grating structures to achieve enhanced absorption (experimental results). (a) Graphene Salisbury screen hybrid metasurface design. The metallic grating between graphene and dielectric layer enhances the interaction between THz waves and graphene. (b) Absorption of the bare graphene and hybrid absorber system at their resonances. (c) Measured reflection as a function of external gate voltage and frequency with the period and width of the grating being 150 um and 70 um, respectively. (d) Absorbance and modulation depth measured at the resonant frequency.[32] Reproduced with permission from Chen *et al.*, Adv. Opt. Mater. 8, 1900660 (2020). Copyright 2020, Wiley online library. (e) Nearly perfect absorber design realized by coupling graphene with 1D PMMA grating. (f) Optical image of the fabricated structure. (g) SEM image of the same structure. (h) Schematic of the measurement setup. (i) The measured and simulated absorption spectra of the hybrid structure with different grating periods for TE polarization.[90] Reproduced with permission from Guo *et al.*, Adv. Opt. Mater. 4, 1955 (2016). Copyright 2016, Wiley online library. (j) Schematic of a perfect absorption structure incorporating graphene ribbons and plasmonic metallic gratings. The graphene plasmonic ribbon is not located in the center of the metallic slit but slightly off the side. (k) SEM image of the structure in (j). (l) Gate voltage-dependent tunable absorption map of the proposed structure in (g).[91] Reproduced with permission from Kim *et al.*, Nano Lett. 18, 971 (2018). Copyright 2018, American Chemical Society.

Table 1. Comparison of representative absorber designs based on hybrid graphene-grating structures.

| Design | Wavelength/ Frequency | Absorption | Bandwidth | Method | References |
|---|---|---|---|---|---|
| Graphene ribbon/Ag grating | 9.6 um | 94% | Single frequency | Simulation | Zhao et al.[35] (2015) |
| PMMA grating/Monolayer graphene/SiO$_2$/Au substrate | 1.5 um | Over 99% | FWHM of 20 nm | Experiment/ Simulation | Guo et al.[90] (2016) |
| Dielectric grating/Spacer/Monolayer graphene | 13.93 THz | 41.97% | Single frequency | Simulation | Wei et al.[66] (2016) |
| Monolayer Graphene/metal/Dielectric /Metal grating | 1.02 um | 90% | Single frequency | Simulation | Pan et al.[33] (2017) |
| Monolayer graphene/Au grating/SiN$_x$ waveguide | 4.5 um | 37% | Single frequency | Simulation | Zhang et al.[82] (2017) |
| Au strips/Graphene ribbons/Salisbury screen | 7.2 um | 96.9% | Single frequency | Experiment/ Theory | Kim et al.[91] (2018) |
| Monolayer graphene/Metal grating | 0.43 THz | 100% | Single frequency | Experiment/ Simulation | Chen et al.[32] (2020) |
| Graphene patch array/Dielectric/ Graphene patch array | 2.4 THz | 99.6% | 1.5 THz | Simulation | Guo et al.[24] (2020) |
| Monolayer graphene/Compound silver grating | 1.67 um | 98% | 0.6 um | Simulation | Nguyen-Huu et al.[85] (2021) |
| Metal grating/Graphene/Planar waveguide | 15.2 um | 76% | Single frequency | Simulation | Ruan et al.[38] (2021) |

## B. Boosting optical nonlinear response

Enhancing the interaction between graphene and incident waves is not only important for linear optics but also even more significant in the nonlinear optics regime, especially at IR and low THz frequencies, where nonlinear effects are usually very weak. Towards this goal, many approaches were reported to strengthen the interaction between graphene and light, such as graphene patterned structures,[13] graphene combined with metal/dielectric nanostructures,[27,92] graphene-based bound states in the continuum (BICs),[93] and the excitation of GPs.[94–96] In addition, graphene itself was proven to be an excellent nonlinear optical material with a large third-order nonlinear optical susceptibility that can potentially enhance various nonlinear effects,[20,97–101] leading to a plethora of new applications in nonlinear optics. Hence, graphene plasmon excitation is an effective method to enhance the nonlinear response of graphene, providing a novel platform to study emerging nonlinear plasmonic effects.

The most attractive advantage of graphene is the gate voltage tunability of its nonlinear conductivity, in a similar and even more sensitive way compared to its linear conductivity. Gate-controllable enhanced third-harmonic generation in single graphene monolayers was experimentally reported over an ultrabroad bandwidth by modulating their Fermi level or tuning the incident photon energy.[102] Significantly increased high harmonic generation efficiency was also demonstrated by probing doped graphene nanostructures, mainly induced by graphene's high intrinsic nonlinearity and the resulted strong plasmonic field confinement.[94] Alternatively, third harmonic generation (THG) strength improvement was demonstrated by exploring the electromagnetic field enhancement caused by a graphene micro-ribbon array.[95] Moreover, the quasi-phase matching technique was investigated to enhance the THG efficiency of propagating SPPs excited on graphene.[103] Graphene patterning led to a further increase in its nonlinear response at the resonant frequency due to the accompanied stronger absorption.[104] Furthermore, the hybridization of graphene with asymmetric plasmonic split ring resonators boosted the nonlinear response compared to bare graphene by more than an order of magnitude within a broad frequency range.[105] In addition, the graphene placement on a grating substrate considerably increased its THG efficiency by several orders of magnitude because of the induced strong coupling effect (see section II for more details). Hence, hybrid graphene-plasmonic grating structures seem to provide an ideal novel platform to increase the efficiency of various nonlinear optical effects.

More specifically, noticeable THG efficiency improvement was reported when a graphene layer was placed on grating substrates.[100] The efficient excitation of THG GPs was demonstrated in graphene sheets again placed on gratings.[12] By attaching two silicon-based waveguides to a graphene-covered dielectric grating, GPs are guided over a wide spectral band at the fundamental frequency (FF), as well as the THG frequency, as depicted in Fig. 5(a). Simulations are performed under normal incident light impinging on the grating section of the dielectric waveguide and Figs. 5(b) and 5(c) present the electric field distributions at FF and THG frequencies, respectively. The FF GPs propagate along the waveguide on both sides after they are excited on the graphene surface exactly above the grating section. Importantly, THG GPs appear on the graphene surface, not only in the grating section, but also along the entire dielectric waveguide system. The total THG conversion efficiency (CE) is calculated for each point along the z-direction and is shown in Fig. 5(d), where its maximum value can reach $3.68 \times 10^{-7}$ at the grating boundary. However, the THG CE value attenuates exponentially along the waveguide due to graphene's Ohmic losses. The maximum THG CE is computed as a function of incident light intensity and the grating period number ($N_p$) with results plotted in Figs. 5(e) and 5(f), respectively. Obviously, the maximum THG CE increases with the power intensity. It can reach $3.68 \times 10^{-7}$ under relatively low input intensity of 0.19 MW/cm$^2$, consisting very low value compared to previous works.[106,107] The simulation results in Fig. 5(f) demonstrate that the maximum THG CE decreases with increased $N_p$. However, the THG output power density is more important to be considered in practical applications, where larger gratings will definitely be required.

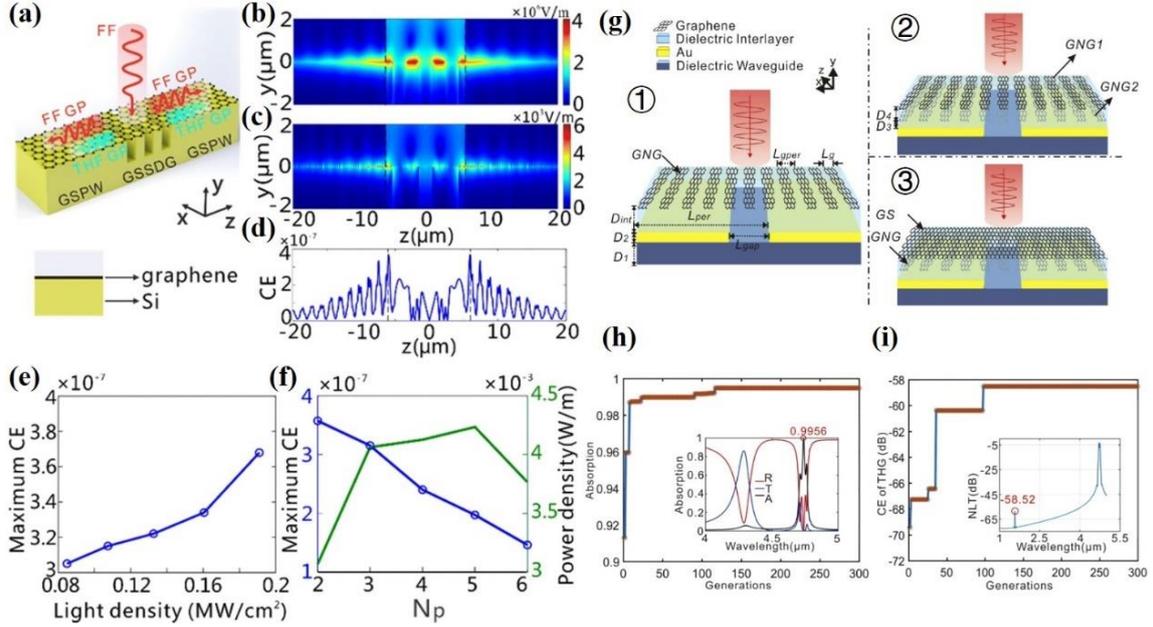

Figure 5. Enhanced nonlinear effects with hybrid graphene-grating designs.
(a) FF and THG GPs based on hybrid dielectric waveguide-gratings covered by graphene. (b)-(c) Electric field distribution at (b) FF and (c) THG frequencies. Only the grating section is illuminated at the resonance wavelength. (d) Computed THG CE along various points at the z-direction. (e) Maximum THG CE as a function of the incident power intensity. (f) Maximum THG CE and power density by varying the grating period.[12] Reproduced with permission from Li *et al*., Nanoscale Res. Lett. 13, 338 (2018). Copyright 2018, Springer US. (g) Three different structures consisting of single- and double-layer graphene. (h)-(i) The evolution of (h) absorption resonant peak and (i) THG CE of the hybrid structure for each GA iteration (called generation). The insert in (h) shows the calculated transmission, reflection, and absorption spectra of the final GA optimized design. The insert in (i) demonstrates the normalized light transmittance of the same design.[108] Reproduced with permission from Yu *et al*., Opt. Express 28, 35561 (2020). Copyright 2020, Optica.

Three different structures are presented in Fig. 5(g) to further enhance the THG CE composed of metallic gratings coupled with patterned graphene metamaterials.[108] The first structure consists of a dielectric interlayer, gold grating, and dielectric waveguide, as well as a graphene nanoribbon grating (GNG) placed on top of the entire hybrid system. In the second design, another GNG (GNG2) is embedded between the gold grating and GNG1. Finally, a third design is presented in Fig. 5(g) that is slightly different from the second, since the upper graphene layer is a monolayer graphene in this case. All these three hybrid structures exhibit a considerably high absorption resonant peak, where the light-matter interaction in graphene is significantly enhanced. Next, genetic algorithms (GA) are used to optimize the designed structures with the goal to further improve the absorption and, consequently, THG CE. Figures 5(h) and 5(i) illustrate the evolution of the absorption peak and THG CE of the hybrid structure after each GA iteration (called generations in the x-axis of these plots), respectively. The maximum absorption of the GA optimized design can reach 0.9956 after 300 generations. Meanwhile, the highest THG CE becomes $1.406 \times 10^{-6}$ under low pump intensity of 0.18 MW/cm$^2$ for the same generation number.

The inherent nonlinear properties of graphene are particularly strong in the THz frequency range.[109] Hence, strong THG was demonstrated in THz owing to graphene's intraband electron transitions leading to its strong interaction with THz radiation.[110] In particular, nonlinear ultrathin metasurface structures were proposed based on periodically patterned graphene micro-ribbons to obtain strong THG at THz frequencies.[20] Extremely high THG CE was achieved with relatively low input intensity values on the order of 0.1 MW/cm$^2$, which makes graphene wave generation more appealing at THz due to the lack of high power sources at these frequencies.[111] As a relevant example, enhanced nonlinear THz effects are reported based on the hybrid graphene-covered metallic grating depicted in Fig. 6(a).[30] Perfect absorption is observed at the resonance of this hybrid grating (Fig. 6(b)) under TM-polarized incident wave indicating a strong coupling effect between the graphene THz plasmon and metallic grating resonance. The left inset in Fig. 6(b) illustrates the equivalent circuit model used to theoretically analyze this structure, where graphene is assumed to be an additional shunt admittance $Y_s$. Note that metallic gratings without graphene can also achieve perfect absorption but lack dynamic tunability combined with strong nonlinearity induced by graphene. The highly localized magnetic fields inside the grating's trenches imply the existence of a magnetic plasmon mode accompanied by strong electric fields that can directly boost the inherent graphene nonlinearity. The right inset in Fig. 6(b) represents the electric field enhancement distribution at resonance. Figure 6(c) demonstrates the computed THG CE by this hybrid grating at its absorption peak as a function of the input intensity. The THG CE is dramatically enhanced with the increase of the incident FF wave intensity. Note that the relationship between the computed THG CE and the input intensity always has a quadratic shape. Importantly, very high THG CE values around 0.16 are achieved under a relatively low intensity of 100 kW/cm$^2$, consisting a substantial improvement compared to previously reported strong THG effects at THz frequencies obtained by relevant nonlinear graphene nanostructures.[20] Furthermore, this work explores another interesting third-order nonlinear optical process, i.e., four wave mixing (FWM), that is also boosted by the same hybrid configuration. The FWM output power of the proposed graphene-covered hybrid metamaterial grating is much higher compared to similar designs but without graphene (only grating), as shown in Fig. 6(d). This result proves that graphene is crucial to obtain strong nonlinearity. Interestingly, the nonlinear response remains relatively insensitive across a broad incident angle range. Thus, the strong enhancement of nonlinear effects achieved theoretically by this simple hybrid configuration can have potential applications in nonlinear THz spectroscopy, THz frequency generators, and wave mixers.

It was computed that graphene has an especially high third-harmonic susceptibility value $|\chi^{(3)}| \approx 10^{-9}\ \mathrm{m^2/V^2}$ at THz frequencies, many orders of magnitude larger compared to near-IR.[112] Based on its strong nonlinearity and easy integration with CMOS-based chips, it is predicted that graphene can be used in chip-integrated THz nonlinear conversion applications.[109] Recently, this concept has been experimentally verified by combining graphene with a metallic grating to provide electric field enhancement at the system's resonance.[109] In this work, the grating-enhanced third-order nonlinearity was explored, where

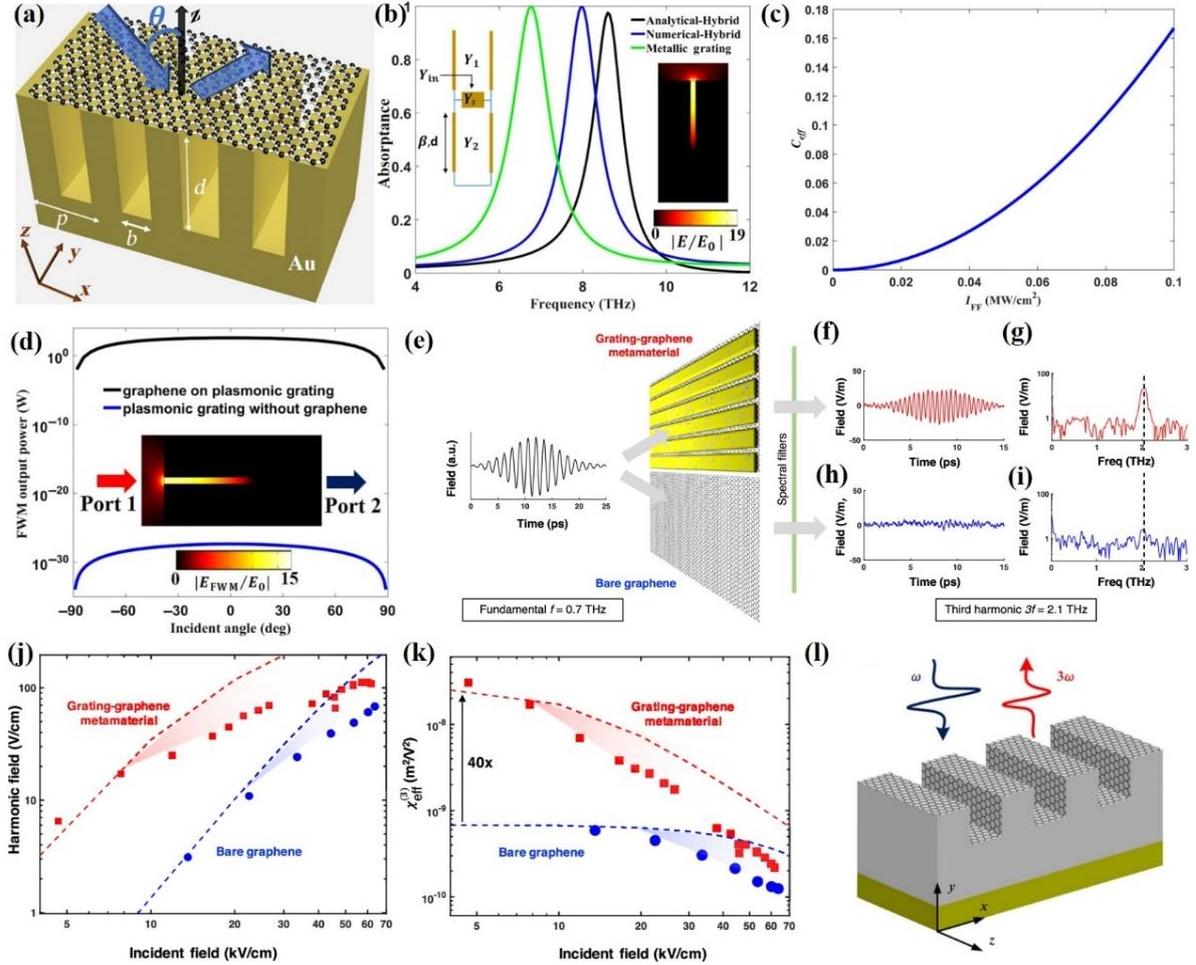

Figure 6. Hybrid grating-graphene metamaterial designs to enhance nonlinear effects.
(a) Illustration of hybrid graphene-plasmonic grating structure. (b) Absorption spectra computed by analytical and numerical methods. The inset on the left is the equivalent-circuit model used to analyze the proposed structure. The inset on the right is the electric field enhancement distribution at the resonance. The Fermi level of graphene is fixed to 0.3 eV. (c) THG CE as a function of the incident intensity for the hybrid configuration. (d) FWM output power as a function of the incident angle $\theta_1$ for the cases of hybrid graphene-grating and grating without graphene presented by black and blue lines, respectively. The inset is the computed electric field enhancement distribution at the FWM frequency.[30] Reproduced with permission from Guo. *et al.*, Phys. Rev. Appl. 11, 024050 (2019). Copyright 2019, American Physical Society. (e) Measurement configuration of a THz wave illuminating the grating-graphene metamaterial (top) and bare graphene monolayer (bottom). (f)-(g) Measured THz fields in (f) time and (g) frequency domain for the grating-graphene metamaterial. (h)-(i) Measured THz fields in (h) time and (i) frequency domain for the bare graphene layer. (j) Measured harmonic fields for the grating-graphene metamaterial and bare graphene layer as a function of incident field strength. (k) Extracted nonlinear susceptibility as a function of incident THz fields for these two cases.[109] Reproduced with permission from Deinert *et al.*, ACS Nano 15, 1145 (2021). Copyright 2021, American Chemical Society. (l) Side view of the graphene-plasmonic grating where graphene covers the entire grating area.[113] Reproduced with permission from Wang *et al.*, Phys. Scr. 97, 115501 (2022). Copyright 2022, IOP Publishing Ltd.

a narrowband THz wave was generated and then focused on the sample. The electric field of the transmitted wave is measured as a function of time by using the configuration shown in Fig. 6(e), where a THz wave with frequency $f = 0.7$ THz illuminates the grating-graphene or bare graphene samples. The signal passes through each sample and is measured in the time domain with results demonstrated in Figs. 6(f) and 6(h), respectively. The transmitted signal has a THG frequency equal to $3f = 2.1$ THz, calculated by using the Fourier transform of its time domain response and depicted in Figs. 6(g) and 6(i) for each sample. Interestingly, no THG signal is observed for the bare graphene case even for higher input field strength of 14 kV/cm, as shown in Figs. 6(h) and 6(i). Hence, the hybrid grating-graphene metamaterial shows a distinct and much stronger nonlinearity. Next, the field strength of the transmitted THG signal at $3f = 2.1$ THz is computed by taking its Fourier transform and extracting the power spectrum peak with results plotted in Fig. 6(j). Clearly, the THG signal of the hybrid grating sample is more than an order of magnitude higher than that of bare graphene. Figure 6(k) illustrates the measured third-order effective nonlinear susceptibility values $\chi_{\text{eff}}^{(3)}$ of these samples that can reach a maximum value of $3 \times 10^{-8}$ m$^2$/V$^2$ and has a negative sign. Interestingly, the nonlinear susceptibility of the grating-graphene hybrid design is more than forty times larger compared to bare graphene. The nonlinear THG CE of the grating-graphene sample can be further improved by increasing the duty cycle and decreasing graphene's Fermi level. The strong saturation effect, depicted in Fig. 6(k) beyond 10 kV/cm, is attributed to the induced high carrier temperatures due to the increased field enhancement. This simple configuration is CMOS-compatible and has low power consumption. Hence, it can be used in commercially viable, chip-integrated, THz nonlinear conversion applications.

Different, but always enhanced, nonlinear effects with hybrid dielectric grating-graphene structures were demonstrated in various previous works, including optical bistability and second harmonic generation (SHG).[12,100,114] The harmonic generation of graphene over silicon gratings was also numerically investigated.[115] The harmonic generation intensities were affected by many factors, such as graphene Fermi level and its carrier mobility, grating period, and incident angle. In most previously proposed hybrid graphene-grating systems, a small part of the graphene area was attached to the grating and interacted with the enhanced fields mainly due to the induced discrete hot spots. To further strengthen the THG CE, a unique design of graphene plasmonic grating is shown in Fig. 6(l), where the entire grating surface, including the trench dips, is conformally covered by graphene layer.[113] The enhanced THG response is expected to be more pronounced by this hybrid system, since it will greatly boost the interaction between incident electric fields and graphene.

**C. Improved modulator and sensor designs**

GPs can be flexibly tuned through optical excitation and electrical gating due to increased chemical doping in graphene.[6,116–118] They also provide a platform to achieve efficient optical modulation in terms of intensity, phase, and polarization of incident electromagnetic waves.[119–121] An absolute phase modulation of 130 ° is experimentally demonstrated with a graphene-based Salisbury screen perfect absorbing device, under the excitation of intense THz pulses.[122] More specifically, the dielectric-graphene plasmonic structures can exhibit nearly 100% resonant transmission when high-mobility graphene is used that can be substantially tuned, a feature that could be very useful to implement THz transmission modulators. Towards this goal, a hybrid graphene/dielectric metasurface system to achieve stronger light-matter interaction was proposed,[27] allowing for improved tunability and modulation functionality. The system is demonstrated in Fig. 7(a). It is composed of periodic asymmetric silicon (Si) nanorods placed between a graphene monolayer and silica substrate. Full-wave simulations are performed by illuminating the structure with a normal incident plane wave. The induced electric field is greatly enhanced and confined around and inside the silicon gratings, since the grating's asymmetry causes the formation of a narrow BIC resonance that leads to high transmission tunability. The transmission coefficients for three different graphene Fermi levels are calculated in Fig. 7(b), where the transmission modulation is clearly demonstrated. The transmission resonance is flatter and broader only in the case of undoped graphene ($E_F = 0$ eV), mainly due to high interband absorption. The graphene's interband transitions loss mechanism is rapidly decreased as the Fermi level increases. As a result, strong transmission modulation is achieved under Fermi levels 0.5 eV and 0.75 eV depicted by the red and blue dashed lines, respectively, in Fig. 7(b). The modulation depth is illustrated more quantitatively as a function of incident wavelength by plotting the absolute value of the transmission difference $\Delta T = |T(E_F = 0.75eV) - T(E_F = 0eV)|$ in Fig. 7(c). Evidently, high transmission modulation, as high as 60%, is obtained at the Fano resonance dip due to the BIC phenomenon. This result indicates efficient and tunable transmission modulation at near-IR frequencies based on BIC metasurfaces loaded with graphene.

Furthermore, two modulators are designed that achieve high quality-factor (Q-factor) resonances with the goal to further increase the modulation depth of hybrid graphene/dielectric metasurface systems.[123] The first design is composed of periodic germanium (Ge) bars adjacent to Ge ring resonators with modulation depth that can reach 93%. This composite metasurface is shown in Fig. 7(d) and consists of two sets of Ge disks with different radii arranged in a certain formation on the silica substrate. Note that a continuous graphene sheet is placed on the Ge disks. Electrons and holes on the top graphene layer are photogenerated by illuminating it with an IR pump leading to two non-equilibrium quasi-Fermi levels $E_{fn} = E_f$ and $E_{fp} = E_f$ created due to the increased intensity of the optical pump. Figures 7(e) and 7(f) represent the transmission and reflection spectra, respectively, for various graphene quasi-Fermi level values. Notable, the transmission dip decreases as the quasi-Fermi level increases. At the same time, the resulted Fano resonance shows a red shift behavior mainly due to Pauli blocking. Hence, the absorption of THz photons is blocked by the optical pumping and carrier photoexcitation. In addition, the reflection peak and Q-factor also increase with the rise in graphene's quasi-Fermi level values. Importantly, the calculated modulation depth can reach as high as 98% at the operating frequency of 5 THz, a much larger value than previously studied graphene-based THz structures. The required optical intensity to achieve this modulation depth is as low as 7.35 W/cm$^2$. This hybrid structure provides a promising potential to design fast and efficient spatial THz modulators.

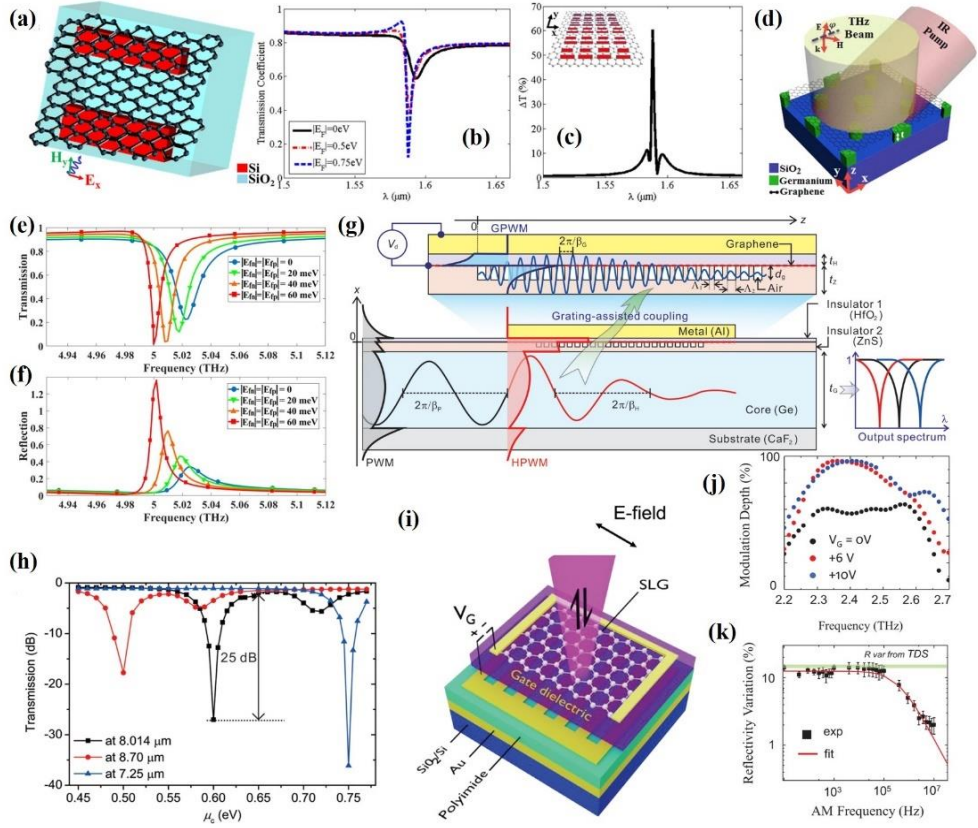

Figure 7. Graphene-plasmon grating applications for transmission or reflection modulation.
(a) Schematic of the hybrid graphene-dielectric grating design to achieve high transmission modulation. (b) Transmission coefficient of the proposed design as a function of incident wavelength under three different graphene Fermi levels. (c) Transmission difference as a function of incident wavelength.[27] Reproduced with permission from Argyropoulos, Opt. Express 23(18), 23787 (2015). Copyright 2015, Optica. (d) Schematic of the proposed all-optical THz modulator. (e)-(f) Calculated (e) transmission and (f) reflection spectra of this structure.[123] Reproduced with permission from Bahadori-Haghighi *et al*., J. Appl. Phys. 128, 044506 (2020). Copyright 2020, AIP Publishing. (g) Schematic and operation principle of a modulator consisting of input and output photonic waveguides. The input photonic waveguide contains graphene on a grating along its slot regions. (h) Transmission spectra as a function of Fermi level $\mu_c$ under three different wavelengths.[124] Reproduced with permission from Kim *et al*., Nanoscale 9, 17429 (2017). Copyright 2017, The Royal Society of Chemistry. (i) Schematic of the proposed modulator, consisting of graphene-covered metal grating on polyimide over gold. (j) Modulation depth versus operating frequency. (k) Modulation speed estimated through reflectance variations.[125] Reproduced with permission from Gaspare *et al*., Adv. Funct. Mater. 31, 2008039 (2021). Copyright 2021, Wiley-VCH GmbH.

Subwavelength modulators are reported in Fig. 7(g) working in the mid-IR frequency range by embedding a hybrid graphene-grating structure into a waveguide.[124] The grating utilized here is essential to achieve efficient coupling between the distinct waveguide modes. The modulation characteristics of this hybrid device are analyzed by changing the driving voltage between the aluminum and graphene layer, leading to different graphene Fermi levels $\mu_c$. The transmission spectra blue shift as $\mu_c$ increases with results demonstrated in Fig. 7(h). The transmission can be tuned by more than 25 dB at wavelength 8 um when $\mu_c$ slightly changes from 0.6 eV to 0.65 eV. In addition, high modulation depth is achieved in every wavelength between 7.25 um and 8.7 um, meaning that the proposed modulator can work in a wide wavelength range. This hybrid design substantially improves the modulation efficiency bandwidth compared to previous related works.

To achieve simultaneously high THz modulation efficiency and speed, graphene-based grating-gated devices were presented[125] that work in reflection mode. The device was fabricated by depositing a polyimide layer on a Au/SiO$_2$/Si substrate, creating a metallic grating. The single layer graphene was grown by chemical vapor deposition (CVD). The relevant schematic of this device is shown in Fig. 7(i). Simulations were performed to model this design and determine the appropriate structure parameters by calculating its reflectance. Then, the modulation performance was measured by a THz time-domain spectroscopy (TDS) system. In order to evaluate the modulation performance, the modulation depth was defined as $\eta = 100 \times [R(V_G) - R(V_{CNP}))/R(V_{CNP})]$, where $R(V_G)$ is the reflectance extracted by the TDS measurements at $V_G$ and $R(V_{CNP})$ is the maximum reflectance at $V_G = V_{CNP}$. Figure 7(j) plots the modulation depth as a function of frequency. It can be seen that $\eta$ can reach a maximum value of ~90% at the resonance frequency of 2.45 THz, while the corresponding insertion loss is low (~1.3 dB). In addition, the modulation speed is measured by placing the designed modulator at the focal point of a quantum cascade laser (QCL) cavity that has single mode emission at 2.68 THz. The modulation speed is estimated through the reflectance variations shown in Fig. 7(k). The 3 dB cutoff frequency obtained from this curve fit ($f_{c.o,fit} = 19.5 \pm 1.8 \,\text{kHz}$) agrees well with the theoretical electronic cutoff $f_{c.o,teo} = 24.5 \,\text{kHz}$.

This work reported a modulator design with an efficient (90%) amplitude modulation with high speed (>20 kHz) that can be integrated with a THz QCL, effectively compensating the QCL cavity dispersion.

Graphene also has high surface-to-volume area ratio combined with large carrier mobility that makes it an attractive material platform for sensing applications. In particular, large sensitivity values and high Q-factors were reported with a graphene-gold grating hybrid sensor.[126] The induced Fano-type transmission or reflection was very sensitive to changes in the surrounding environment of this structure. Small changes in the refractive index caused a significant shift in the resonance wavelength, achieving highly sensitive detection. Moreover, tunable and ultracompact refractive index sensors were constructed by using the metal-graphene hybrid nanostructures shown in Fig. 8(a).[42] In this case, both graphene surface plasmon (GSP) and waveguide resonant modes in the silver grating slits are generated based on the alternative deposition of graphene and silver nanoribbons on the metallic substrate. This leads to a Fano resonance, originating from the coupling between the GSP and grating waveguide mode. Figure 8(b) illustrates the simulated spectra under various ambient refractive indices. The sharp Fano transmission dip shows a redshift as the refractive index values are increased, consisting a crucial property to obtain high sensitivity when refractive index variations in the surrounding environment are present. Both Fano transmission dip and peak exhibit a blueshift as the Fermi level increases with results depicted in Fig. 8(c). The GSP electrical tuning is the signature feature of graphene-based devices, usually leading to a broadening in the operation bandwidth of the designed structure. Variations of the structures were also studied to investigate the fabrication tolerance of these hybrid gratings. The thickness of the substrate, middle silver film, and grating period had little effect on the sensing performance.

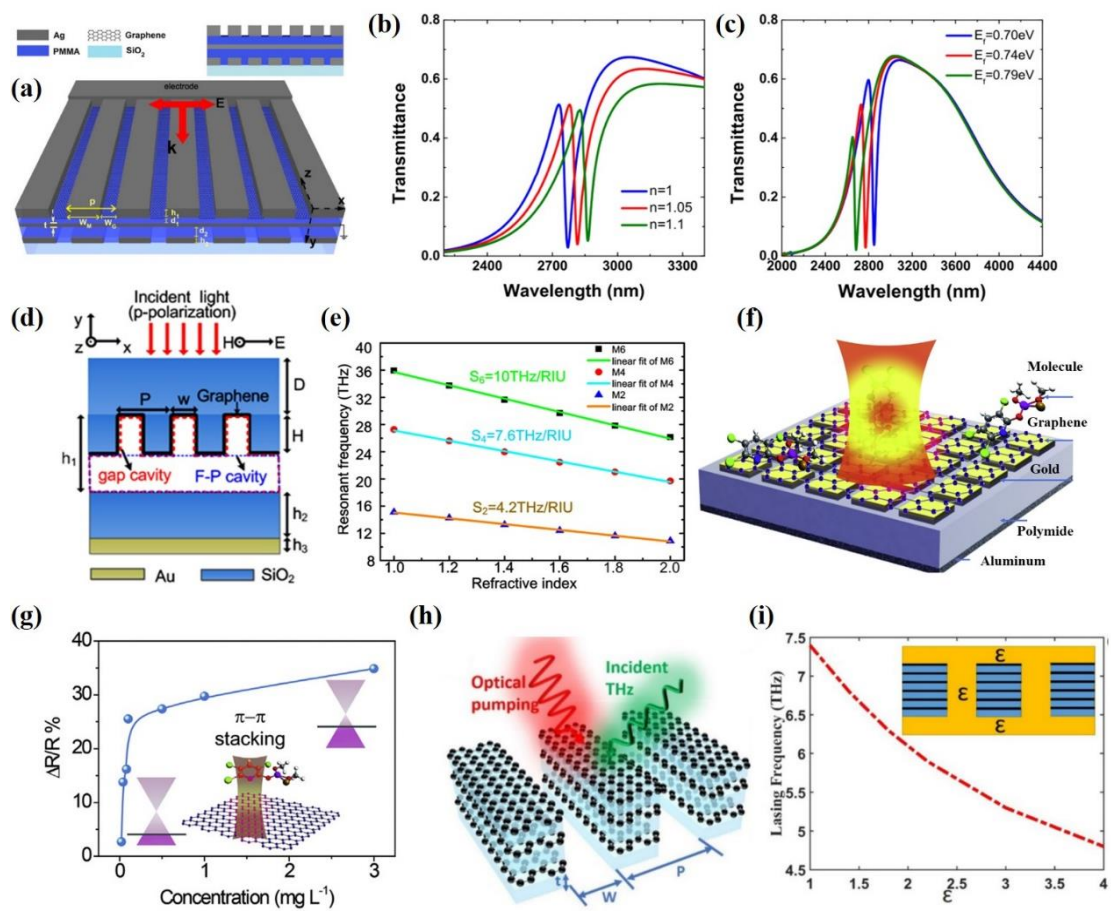

Figure 8. Graphene-plasmon grating applications in sensing.
(a) Schematic of the proposed silver-graphene hybrid structure to achieve a refractive index sensor. (b) Transmission versus operating wavelength under different ambient refractive indices. (c) Transmission versus wavelength under different graphene Fermi level values.[42] Reproduced with permission from Pan *et al.*, Sci. Rep. 6, 1 (2016). Copyright 2016, Nature Publishing Group. (d) The front view of the proposed ultrasensitive THz sensor. (e) Resonant frequency versus refractive indices of the analytes to evaluate the sensing performance.[39] Reproduced with permission from Yan *et al*., J. Light. Technol. 37, 1103 (2019). Copyright 2019, IEEE. (f) Schematic of the designed sensor based on monolayer graphene-coated THz metamaterials. (g) Relative reflectance changes when adding chlorpyrifos methyl molecules on top of the graphene-metamaterial heterostructure with different concentrations ranging from 0.02 mg/L to 3.0 mg/L.[127] Reproduced with permission from Xu et al., Carbon N. Y. 141, 247 (2019). Copyright 2019, Elsevier Ltd. (h) Schematic of the proposed hyperbolic metamaterial structure with high sensitivity to the surrounding medium. (i) Lasing frequency versus dielectric permittivity when the proposed structure is embedded in different dielectric materials.[62] Reproduced with permission from Guo *et al*., Opt. Mater. Express 8, 3941 (2018). Copyright 2018, Optica.

Single monolayer graphene gratings were integrated with Fabry-Perot (FP) cavities to further improve the sensitivity in the THz frequency regime.[39] Figures 8(d) represents the front view of the proposed system. The incident field was confined in a microscale cavity leading to higher sensitivity. The sensing performance was evaluated by performing full-wave simulations with results plotted in Fig. 8(e), where a frequency shift in all resonant modes is obtained by changing the analyte refractive index values. The slope of the fitting lines is utilized to indicate the frequency sensitivity. Here, peak levels of 4.2, 7.6, and 10 THz/RIU are achieved at the M2, M4, and M6 modes, respectively. In the same work, the tunability and wide-angle sensing operation were illustrated by calculating the absorption for various graphene Fermi levels and under oblique incidences, respectively. Based on these advantages, the proposed graphene-based scheme is promising for the design of on-chip integrated THz biochemical sensors.

The first experimental demonstration of a THz sensor based on hybrid graphene-metamaterial heterostructure was realized by adding chlorpyrifos methyl molecules on top of this structure.[127] The schematic of the proposed heterostructure sensor is shown in Fig. 8(f), which is composed of monolayer CVD-grown graphene, planar array of square patch gold resonators, thin layer of polymide, and aluminum ground plane. The reflection spectra are measured by using a THz time-domain spectrometer system, defined as $R = |E_{sample}/E_{reference}|^2$ with $E_{sample}$ ($E_{reference}$) the THz electric field intensity reflected from the sample (silver mirror reference). The relative reflectance change is expressed as $\Delta R/R = (R_{reference} - R_{sample})/R_{reference}$, where $R_{sample}$ ($R_{reference}$) is the measured reflectance with (without) chlorpyrifos methyl. Figure 8(g) illustrates the relative reflectance changes with chlorpyrifos methyl solution added on top of the heterostructure. It is found that the value $\Delta R/R$ rapidly increases when the concentration is lower than 0.1 mg/L, while it rises slowly with the growth of the concentration. It can be detected that a maximum $\Delta R/R$ of 35% is obtained when the concentration is 3.0 mg/L. This high sensitivity originates from the π-π stacking forming between molecules and graphene. If the structure is further optimized, the sensitivity can be greatly improved, leading to a plethora of new THz sensing applications.

Likewise, a novel physical mechanism was explored to enhance the sensitivity of graphene-based devices by designing hyperbolic metamaterial (HMM) structures made of photoexcited graphene and dielectric layers that were stacked in an alternative way.[62] Figure 8(h) illustrates the proposed structure, where the active graphene response is achieved due to IR/visible light pumping. Graphene can act as a gain medium under sufficiently strong photoexcitation levels due to the resulted population inversion.[128,129] The light-trapping effect was supported by this HMM design at THz frequencies. Hence, the THz gain of the photoexcited graphene was boosted leading to lasing effect. Simulations and theoretical analysis were performed to investigate the lasing or amplification response. It was found that this response is ultrasensitive to the surrounding media when the HMM is embedded in different dielectric materials, as demonstrated in the inset of Fig. 8(i). The substantial redshift in the THz resonant frequency of the lasing mode is presented in Fig. 8(i) achieved by changing the surrounding dielectric medium. This result indicates strong THz sensing.

**D. Efficient photodetectors**

Photodetection consists a very important property for a range of fields, such as optical communications, environmental monitoring, and image sensing. The first example of photodetectors based on graphene exhibited photo-responsivity of 6.1 mA/W at 1.55 um.[41] This work combined graphene with plasmonic nanoparticles to enhance the responsivity of hybrid photodetectors. However, its photoactive region was limited to submicron length scale because of the metal-graphene junction small area. Hence, recently, SPPs graphene photodetectors were proposed to extend the junction region by coupling graphene with metal gratings,[53] as shown in Fig. 9(a). The photovoltage maps were plotted at different wavelengths based on this optimized geometry. Figure 9(b) compares the structured grating contact and the flat contact at 785 nm, where a large enhancement of photoresponse (~400%) is achieved. The grating area determines the photoactive area, since the light energy delivered to the graphene junction occurs at the contact edges where light is converted to an electrical signal. The physical mechanism behind the response in this work was the coupling of the incident light into SPPs, which provides an interesting platform to achieve SPP-based photodetectors by using hybrid metallic gratings.

Interestingly, a novel mechanism of photoconductive gain induced by plasma waves was reported in a graphene field-effect transistor (FET) design.[54] Grating-gates were utilized to excite the plasma waves under THz excitation. The excited plasma waves oscillated in the graphene FET channel, serving as a cavity, with schematic shown in Fig. 9(c). The transfer-characteristic and induced photocurrent of the proposed graphene FET are presented in Figs. 9(d) and 9(e), respectively. The photoresponse was enhanced by increasing the drain bias voltage. The conductance tunability was realized in this work by changing the incident frequency, AC voltage-source configuration, or direct-current driven electric fields. Metal nanogratings integrated with monolayer graphene were utilized to achieve even higher performance graphene-based photodetectors.[52] The 3D schematic illustration of this design is shown in Fig. 9(f), where SPPs are excited under TM polarization near the metal-graphene interface leading to enhanced absorption. With the generation of electron-hole pairs in graphene, a built-in electric field separated the photoexcited carriers. As a result, photocurrent and electrical signals emerged. The optimal values of the structure's parameters were determined by using simulations with the goal to obtain enhanced absorption to generate maximum carriers. Maximum photocurrent of 0.81 mA at 3 V bias voltage was obtained by computing the I-V curve of this structure response under an incident power $P$ of only 0.5 mW. The responsivity is plotted as a function of bias voltage in Fig. 9(g). It can be observed that a high responsivity of 1650 mA/W is achieved by just using 3 V bias voltage. In addition, the relationship between the responsivity and working wavelength is also investigated under different bias voltages and presented in Fig. 9(h). The increase in bias voltage values resulted in spectral response broadening because of the rise in electric field values.

Next, structure asymmetry was introduced in hybrid graphene-grating designs to couple the incoming radiation more efficiently to the FET channel. In the case of THz photodetectors, the photovoltaic response was enhanced only when asymmetry was introduced into the FET unit cell design.[56,130] As an example, asymmetric grating gates were used to improve the responsivity of graphene-based FET detectors shown in Fig. 9(i).[31] In this work, two flakes of h-BN were utilized to completely encapsulate the monolayer graphene ensuring its high quality, as depicted in Fig. 9(i). Larger photocurrent existed for 0.3 THz compared to lower frequencies, where a higher signal-to-noise ratio was present. This can be explained due to the stronger radiation coupling to the FET structure at higher frequencies. Maximum current responsivity $R_i$ with values up to 55.6 mA/W is obtained at cryogenic temperatures, as demonstrated in Fig. 9(j). Applications in THz sensing and inspection of hidden metallic objects are envisioned based on these new graphene-based FET detectors.

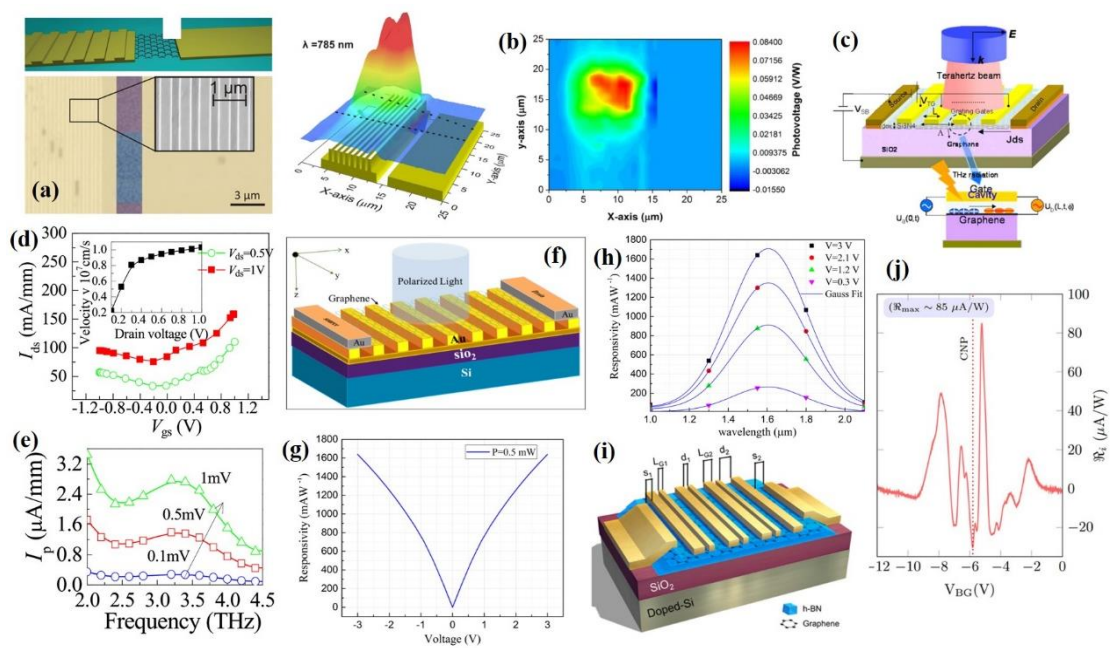

Figure 9. Hybrid graphene-metal gratings to enhance photodetection.
(a) Schematic and SEM image of the proposed metal-graphene-metal photodetector. (b) Scanning photovoltage maps under 785 nm excitation.[53] Reproduced with permission from Echtermeyer *et al*., Nano Lett. 16, 8 (2016). Copyright 2016, American Chemical Society. (c) Schematic of the hybrid graphene-grating structure to realize photo-conductance. Here, two AC voltage sources are used to excite graphene simultaneously. (d) Photocurrent characteristics of the proposed design as a function of bias voltages $V_{gs}$. (e) Photocurrent corresponding to the fundamental plasma resonance at 3 THz.[54] Reproduced with permission from Wang *et al*., Nanotechnology 27, 035205 (2016). Copyright 2016, IOP Publishing Ltd. (f) Schematic of an alternative graphene-based photodetector. (g) Responsivity versus bias voltage at 1.55 um wavelength. (h) Responsivity versus wavelength under different bias voltage values.[52] Reproduced with permission from Khosravian *et al*., J. Opt. Soc. Am. B 38, 1192 (2021). Copyright 2021, Optica. (i) Asymmetric grating-graphene design to achieve gate-bias-dependent photocurrent. (j) Effective current responsivity versus the bias voltage when the excitation is fixed to 0.3 THz.[31] Reproduced with permission from Delgado-Notario *et al*., APL Photonics 5, 066102 (2020). Copyright 2020, AIP Publishing.

### E. Nonreciprocal structures

The induced electric field distributions can be different for opposite excitation directions due to the structure's asymmetry.[131–134] When nonlinearities are considered in these asymmetric structures, nonreciprocal phenomena are realized. These interesting effects can also be realized by asymmetric hybrid grating designs combined with graphene. As an example, the nonreciprocal transmission in near-IR frequencies of such asymmetric HGPG is demonstrated in Fig. 10(a), where designs of nonlinear monolayer graphene integrated with low-loss dielectric gratings are theoretically presented.[135] In the linear regime, the maximum electric field enhancement is found at the resonant wavelength and plotted in Fig. 10(b). Then, the optical nonlinear Kerr effect is triggered under strong input laser excitation intensities. The change in the effective refractive index of the dielectric and graphene, mainly due to Kerr effect, shifts the resonance peak and affects the transmission spectrum, as shown in Fig. 10(c). However, the induced electric field maximum values are different in the case of either forward or backward illumination mainly due to the structure's geometrical asymmetry. As a result, the refractive index changes differently for opposite direction illumination, achieving nonreciprocal transmission which is demonstrated in Fig. 10(d). This work presents a relatively simple HGPG design to realize self-induced nonreciprocity due to nonlinearities paving the way to achieve other relevant nonlinear phenomena, such as optical bistability. The nonreciprocal efficiency can be further improved by increasing the HGPG structure's asymmetry. Moreover, this type of design can work in a broad frequency range just by appropriately tuning the grating parameters.

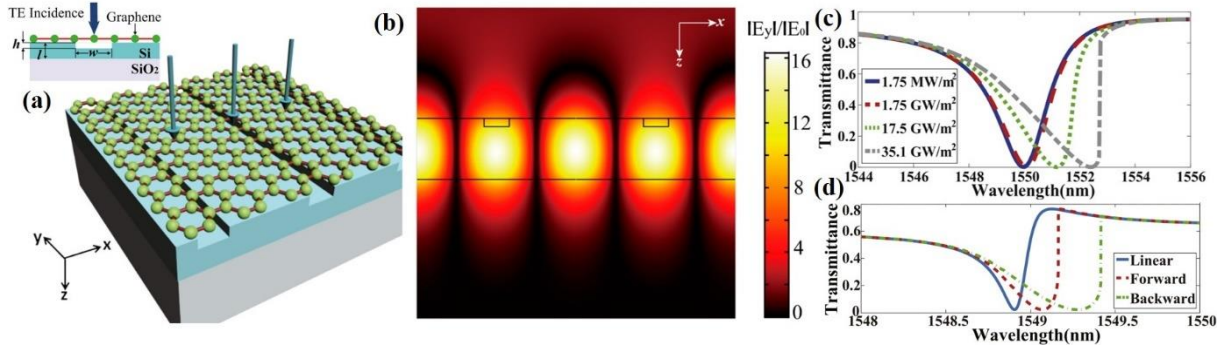

Figure 10. Hybrid graphene-dielectric gratings to achieve nonreciprocal transmission.
(a) Schematic of the nonlinear metasurface based on monolayer graphene integrated with asymmetric dielectric grating. (b) Electric field enhancement at the resonant wavelength in the linear regime. (c) Transmission in the nonlinear regime as a function of wavelength under different incident intensity values. (d) Nonreciprocal transmission as a function of wavelength in the case of forward or backward laser illumination.[135] Reproduced with permission from Chen *et al.*, Carbon N. Y. 173, 126 (2021). Copyright 2021, Elsevier Ltd.

# IV. SUMMARY AND PERSPECTIVE

HGPGs are a promising class of emerging structures exhibiting unique properties that are not present in either graphene or grating alone. This perspective paper illustrated how the HGPG designs enable superior performance compared to other relevant designs based solely on graphene or gratings. We emphasize some unique features of GPs, such as high tunability, strong light confinement, and low loss compared to conventional metal-based plasmons. Optical gratings provide strong localized electric fields and are widely used to excite GPs more efficiently along a continuous graphene layer. Hybrid designs consisting of grating structures combined with graphene benefit from the merits of both systems. More importantly, strong coupling between GPs and gratings plays an important role in improving the performance of these complex structures. Various applications of HGPG in diverse research areas are presented, such as perfect absorbers, nonlinear structures, modulators, sensors, photodetectors, and nonreciprocal devices. The physical mechanisms are highlighted for each application by demonstrating theoretical analysis and experimental implementations.

Although the recent progress of graphene-based hybrid structures has shown great opportunity in various applications, several challenges still exist that prohibit the establishment of this new technology. First, the graphene simulation models need to be further improved to consider more practical parameters. For example, we can introduce additional parameters to precisely define graphene material properties, since the surface conductivity model used in the majority of current simulations is valid only for single graphene layer. Second, large electron momentum relaxation time is extremely challenging to be achieved in experiments involving the more practical large-area CVD-grown graphene[136]. It should be noticed that the relaxation time value used in numerical simulations are usually taken one or even two orders of magnitude higher than that in practical experiments,[137] because ideal conditions are assumed in the fabrication process mainly in terms of temperature, doping level, external field, and various substrate materials, to name a few. It has been demonstrated that larger relaxation time will introduce stronger absorption. As a result, improving the quality of graphene used in experiments will be a significant milestone to boost the efficiency of HGPG structures. Furthermore, the shift in the plasmonic resonance frequency of HGPGs to near-IR or even visible frequency bands still remains challenging. Fortunately, some progress has been recently reported in this field, for example, higher doping to increase the carrier density through optical pumping or electron injection.[137] By combining patterned graphene arrays with dielectric/metallic layers or plasmonic nanostructures, significant enhanced absorption of graphene can be realized at smaller wavelgnths.[22]

Very recently, the concept of BIC has emerged in photonic systems,[138,139] offering extremely high Q-factor resonances combined with enhanced electric fields. We envision that the utilization of the BICs effect generated by HGPG systems will further boost the efficiency of nonlinear optical phenomena and other relevant effects requiring strong light-matter interaction. In addition, BIC phenomena are highly sensitive to the surrounding environment, which is expected to result in useful hybrid designs for biosensing applications. Moreover, the nonequilibrium relaxation dynamics of photoexcited graphene is another interesting emerging research topic that can be combined with gratings and lead to novel applications.[129,140] Negative dynamic conductivity can be realized in photoexcited graphene, leading to lasing or amplification effects. Efficient tunable lasers can be realized by incorporating photoexcited graphene into grating structures, operating across a broad range of wavelengths. Furthermore, the integration of graphene with other novel nanomaterials, such as quantum dots or even alternatively 2D materials (or van der Waals heterostructures), can lead to the discovery of new phenomena and enable us to explore the fundamental limits of light-matter interaction at the nanoscale. In the future, we expect significant advancements in the fabrication techniques and characterization methods of hybrid graphene structures, paving the way for the development of high-performance photonic and optoelectronic devices, such as biosensors, lasers, solar cells, and ultra-compact photonic circuits.

# ACKNOWLEDGMENTS

T. G. acknowledges support from the National Natural Science Foundation of China (NSFC) (Grant No. 12104203). C. A. acknowledges partial support from the National Science Foundation under Grant No. 2224456 and 2212050, and the Office of Naval Research Young Investigator Program (ONR YIP) (Grant No. N00014-19-1-2384).

# DATA AVAILABILITY

The data that support the findings of this study are available from the corresponding author upon reasonable request.